Title: **Its All on the Square- The Importance of the Sum of Squares and Making the General Linear Model Simple**

Alexander Nussbaum
St John's University
Analytic Medtek Consultants LLC
Richard Seides
Seton Hall University

Statistics is one of the most valuable of disciplines. Science is based on proof and it alone produces results, other approaches are not, and do not. Statistics is the only acceptable language of proof in science. Yet statistics is difficult to understand for a large percentage of those who will be evaluating and even doing research. Reasons for this difficulty may be that statistics operates counter to the way people think, as well as the widespread phobia of numeracy. Adding to the difficulty is that undergraduate textbooks tend to make statistical tests seem to be an unorganized conglomeration of unrelated procedures, and this leads to a failure of students to understand that all of the parametric procedures they are studying in an introductory course are ultimately doing the same thing and stem from common sources. In statistics, precisely because the material is complex, the presentation must be simple! This article endeavors to do just that.

### The Problem in Teaching Statistics

Statistics is among the most valuable subjects for anyone to learn but unfortunately, it is difficult to teach. It requires a probabilistic method of thinking that is different from that in other courses -- even other math courses. Science produces knowledge about empirical reality, and statistics is the language of proof in science. As put by the father of statistics, Sir Francis Galton (1889), "Some people hate the very name of statistics, but I find them full of beauty and interest…They are the only tools by which an opening can be cut through the formidable thicket of difficulties that bars the path of those who pursue the Science of man." Over 120 years later, statistics are still universally seen as the basis for judging whether data support a hypothesis. Without statistics it is impossible to judge the efficacy of medical and other procedures, and without a knowledge of statistics it is impossible to fully understand the literature in psychology and many other disciplines. Even those who are only consumers rather than producers of research will face the problem of failure of replicability of medical research as well as the more general problem of whether the data presented demonstrate an article's claims. Not understanding statistics is entering a battle unarmed.

But statistics is among the hardest subjects to comprehend. Anything with numbers intimidates students. Statistics may even be counter to how people usually think! Carl Sagan (1995) noted that even President Eisenhower was alarmed upon learning that one-half of all Americans were of below average intelligence!

Statistics is so substantial and valuable a field, yet so daunting, that the essential task in teaching it is maintaining its integrity while in every-way possible aiming for ease. It is no feat and of no merit to convince students of the complexity and difficulty of statistics. It is however both a feat and a worthwhile undertaking to try to convince students that statistics is comprehensible and even, as Galton felt, beautiful. There is a big difference between complexity of material and simplicity of presentation. Only a presentation that strives to *simplify* statistics to the greatest extent possible, and maximizes the use of everyday language and vivid examples, can possibly succeed in explaining the *complexity* of statistics to a group of tentative students. The situation statistics faces in common with science is well illustrated by a quote by computer scientist and musician Jaron Lanier (Brockman, 2003):

> The arts and humanities have been perpetually faced with the challenge of making simple things complicated. So there exist preposterously garbled academic books about philosophy and art....Science faces the opposite problem. (P.368)

**Parametric statistics and the concept of the Sum-of-Squares**

What we want to do here is to provide a simple explanation for what the parametric statistical procedures the undergraduate students will be taught are doing, and also emphasize the underlying identicalness of all these procedures.  Simply put, parametric statistics  assume a shape for the population distribution, generally a normal distribution, and makes inferences about the measures of that distribution, such as the mean.  We believe that  the key  to presenting  introductory statistics  in its simplest and most comprehensible form is an approach that  emphasizes the importance and understanding of the idea of the Sum of Squares is [1]. Not only is this not currently commonly done, some textbooks go as far as presenting the definitional formula of sum of squares when going over the standard deviation, and then shift to a computational formulas when doing ANOVA or correlation [2].  Students thus fail to grasp that we are still computing the same thing, the  Sum of Squares.

When students understand what Sum of Squares is doing, they will better understand various statistical procedures.  In most parametric statistics the total variability (how much scores vary from one another) is calculated and the concern is determining what percentage is due to the independent variable. All this is based on the general linear model utilizing the method of least squares. Of course instructors understand this. But this needs to be communicated to students in the simplest possible way, and is usually not. Certainly textbooks do not make this clear [3].

**The Sum of Squares**

The sum of squares, short for the sum of squared deviations from the mean, is omnipresent throughout statistics. Statistics is all about variability, how much numbers vary from one another. Our basic measure of variability is called the "variance," which is sum of squares divided by the number of scores that are free to vary, namely N, the total number of scores, in a population and n-1. i.e., degrees of freedom, in a sample. The concept of degrees of freedom will be explained below.

Undergraduate students, graduate students, and even some faculty members have a hard time understanding statistics. Early in a statistics course, the critical concepts of Sum of Squares, variance, and standard deviation pose difficulty.

For our basic measure of variability we want a gauge of average differences from the mean. Thus, we subtract the mean from each score. It may at first blush seem we could merely add up these differences and divide by how many scores, however the sum of deviations from the mean is always zero. We get around this by squaring each deviation, hence the sum of squares [4].

The definitional formula for the Sum of Squares is:

$$\sum_{i=1}^{n} \left(X_i - \overline{X}\right)^2$$

This indicates subtracting the mean from every score, squaring the result and then adding all the squared deviations from the mean. The sample mean is symbolized xbar, ($\bar{x}$). The i=1 means you start summing with the first score, the n being the total sample size, and you continue summing until the last score. When more data is added, the sum of squares will generally increase simply because there are now more scores that can differ from the mean. So we divide by degrees of freedom giving us the variance. The idea of degrees of freedom is rather simple. If you are told a mean is based on 3 scores and given the mean, there is no way for you to know what the scores are. However if you are told two of the scores, the third is locked in; you therefore know it. Likewise, if a mean is based on ten scores and you are given nine, the last one is known. Thus with n scores and a mean, n-1 are free to vary, the last is locked in place. A degree of freedom is lost when the sample mean, which is only an estimate of the population mean, must be used to calculate sum of squares [5].

While we account for how many scores could have produced variability by dividing by degrees of freedom, it is important to understand that the Sum of Squares by itself is a measure of variability. Textbooks generally do not seem to make the above clear, which prevents using the Sum of Squares to demonstrate the unity of all parametric statistics covered in undergraduate course. A close analogy is hits in baseball. Number of hits by itself is a measure of batting success, however we divide number of hits by number of opportunities to get batting average  That is exactly the relationship  Sum of Squares has to variance.

**History**

   The Sum of Squares is essential to Standard Deviation, and understanding the normal distribution implies understanding the latter. By 1733 Abraham de Moivre had described the normal distribution (Hald, 2003) which meant that he understood the parameter of standard deviation, though not naming it as such (McGarth, 1999). The discovery and first use of the Method of Least Squares, which is at the heart of regression, turned into a dispute in priority between Carl Gauss and Adrien-Marie Legendre (Stigler 1981). Legendre published the method in 1805; Gauss had a claim to have discovered it in 1794.  The methods of least squares is the technique of minimizing the sum of the squared errors. In correlation for example, the regression line, our best formula for predicting one variable from another,  is  the line that minimizes the squared distances of all the data points to the line. This minimizes the sum of the squared  actual minus predicted scores.  In any case, this concept utilizing Sum of Squares is just over 200 years old. The term "standard deviation" was coined by Karl Pearson in 1893 (Gillispie, 1970) although as we just saw the idea is much older.  Concepts very close to the current "Standard deviation" have  been referred to as "error of the mean square" or "mean error"  since the early 19$^{th}$ century,  Carl Gauss using the latter term in 1821 (Sprott, 1978).

**Partitioning the Sum of Squares-What does it Mean?**

Endeavored below is a simple way of explaining to introductory statistics students what we are doing in an ANOVA. Say we have four scores: 11, 7, 30 and 20. By how much do they vary? We want to know how much each score differs from the mean, this seems an intuitive way to get variability -- so we subtract the mean from every score. We find that the sum of deviations from the mean is always zero (this may not be immediately intuitive but is a consequence of what the mean, or average, actually is). So we square each deviation and add them together. We now have the sum of squared deviations from the mean or the Sum of Squares. Obviously, at this point we could divide the Sum of Squares by N or n-1 in order to get the variance and then take the square root to get the standard deviation. However, for simplicity's sake, only the Sum of Squares as the measure of variability will be discussed. Consider our limited data set below:

$(11-17)^2 = \phantom{0}36$

$(07-17)^2 = 100$

$(30-17)^2 = 169$

$\underline{(20-17)^2 = \phantom{00}9}$

$\phantom{00000}\Sigma = 314$

Our Sum of Squares is 314. This is the measure of total variability in our four scores.

Let us say, however, that these four scores came from two groups, the scores in group one were 11 and 7 and the scores in group two were 30 and 20. This can represent a very small pilot study.  For example, two participants were placed in a high noise condition (group one) and two participants were placed in a low noise condition (group two).  The scores are some measure of performance on a task. How much do the 11 and 7 differ by?

$(11-9)^2 = 4$

$(7-9)^2 = 4$

The Sum of Squares group one is 8 (4 + 4).

By how much do the 30 and 20 differ? The Sum of squares of group two is 50. To illustrate: $(30 – 25)^2 = 25$, $(20 – 25)^2 = 25$; $25 + 25 = 50$.

The Sum of Squares within both groups combined is 58. (Note we are now using the within groups mean not the overall or grand mean.) Why do the scores in group one differ from one another? Why do the scores in group two differ from one another?  The answer can be explained by subject (participant) differences and/or error in measurement). Think about what we have said so far. The total variability is 314.  The variability due to subject differences is 58 , which means the variability due to group (due to being in a low noise versus high noise environment, in our example) is 314-58, which equals 256 SS.

Let us look at the ratio of variability due to the group (the treatment) to the total variability, which is the total SS. This is simply all we will be doing in regression analysis or Analysis of variance, which are the same thing .  We have just covered the underlying logic of parametric statistics.

The ratio of the variability due to treatment (high noise vs. low noise) over all the variability is 256/314 or .815. We call this number $r^2$ or the coefficient of determination.

Therefore, the correlation coefficient ("r"), is the square root of .815 or .90. If we wish to predict scores just from knowing what group they are in the model will account for 81.5 % of the variability in scores. In our case we have simple regression as we have one predictor variable, the group, and one dependent variable, the score. This accounting of variability is symbolized in regression as $R^2$.

By partitioning the Sum of Squares, it is now easy to perform an Analysis of Variance (ANOVA). We can enter the following into our ANOVA table: SS between = 256, SS within = 58 and SS total = 314. With two groups df between is going to be 1 (it is groups -1), df within is $n_1-1 + n_2-1$, here 1+1=2. Between groups represents differences in scores (variability) due to what group they came from and within groups represents differences in scores (variability) due to subject differences.

Our full ANOVA table is:

|  | Sum of Squares | df | Mean Square | F | Sig. |
|---|---|---|---|---|---|
| Between Groups | 256.000 | 1 | 256.000 | 8.828 | .097 |
| Within Groups | 58.000 | 2 | 29.000 |  |  |
| Total | 314.000 | 3 |  |  |  |

T-tests are a special case of ANOVA with just two means, $t^2 = F$. If this was treated as an independent groups design for which a t was computed, t would equal 2.97. Analysis of variance can be viewed as a special case of regression in which the predicting variables are categories rather than continuous.

There is one additional thing to consider. In our example, we were analyzing a true experiment, so had we found significance, we would therefore be entitled to say that high levels of noise reduced performance. But we could have had the exact same data with nature assigning the group. In which case regardless of whether the results are expressed in F or r, we cannot conclude cause and effect.

The sum of squares is used in T-tests, ANOVA, correlation and multiple regression which are all related. It is thus a central concept in just about all the inferential statistics undergraduates will encounter.

Notes :

[1]The authors recognize that there have been techniques developed for handling parametric data which do not involve sum of squares, however these are generally not covered in undergraduate statistics courses. There are methods of parameter estimation such as maximum likelihood estimation that do not employ the least squares method, however in addition to being beyond what undergraduates can reasonably be expected to encounter in a statistics course, the two methods obtain similar results if the assumption of normality is met.

[2] Before the days of calculators the computational formula was somewhat easier to use, because it did not require subtracting mean from each score, however though of

course algebraically equivalent to the definitional formula, it obscures what Sum of Squares is .

[3] Obviously we will present only the simplest case of linear regression below.

[4] It should be apparent that another strategy could have very well been used, namely, taking the absolute deviation from the mean, adding up the absolute deviations, and dividing by scores to obtain the mean absolute deviation.

Whether standard deviation (SD) or mean absolute deviation (MAD) should be used turned into a debate in the years 1914-1921 between two science heavyweights, astrophysicist Sir Arthur Eddington and statistical pioneer and geneticist Sir Ronald Fisher. Eddington advocated MAD while Fisher showed that SD more efficient with normal distributions and his position carried the day (Gorard, 2005).

However Tukey (1960) confirmed what Eddington had claimed, that MAD was better in distributions which deviate even slightly from normal and obviously in the real world, even the most normal of distributions deviate from normal. Basically when drawing numerous samples from a normal distribution the standard deviation of the MADs is higher than the SD of the standard deviations, however, the situation reverses when drawing form a distribution that departs even slightly from normal. Regardless, other advantages can be found for SD, which is here to stay. A factor in its victory was that using the concept of absolute values makes calculations involving variability more complicated than simply uses squaring. The choice can be ultimately viewed on some level as arbitrary, and in any event the MAD of a normal distribution is .7979 times the SD (see Vince, 2007 p. 58-59 for an explanation).

[5] Another way of understanding the necessity of making the n-1 adjustment when calculating sample variance is as that by using the sample mean we are sure that the resultant sum of squares will be smaller than would have resulted had we known and used the population mean. The sample mean is either larger than population mean or smaller, either way the SS gotten using it smaller than true SS. How much smaller depends on how close sample mean is to population mean. The closer it is, the closer SS using the sample mean is to the correct SS using population mean. The larger our sample the closer the sample mean is likely to be to population mean. So a smaller adjustment to denominator of standard deviation formula need be made, and by using n-1indeed a smaller adjustment is is being made. Dividing by 4 instead of 5 makes a noticeable difference, dividing by 999 instead of 1000 makes no practical difference.